\documentstyle[12pt,aps]{revtex}

\begin{document}

\title{Supersymmetry and Large Scale Left-Right Symmetry}
\author{Charanjit S. Aulakh$^{(1)}$, Alejandra Melfo$^{(2)}$, 
Andrija Ra\v{s}in$^{(3)}$ and Goran Senjanovi\'c$^{(3)}$} 
\address{$^{(1)}$ {\it Dept. of Physics, Panjab University,
 Chandigarh, India}\\
$^{(2)}$ {\it International School for Advanced Studies,
 Trieste, Italy, and CAT, Universidad de Los Andes, M\'erida, 
Venezuela}\\
$^{(3)}${\it International Center for Theoretical Physics, 
34100 Trieste, Italy }}

\maketitle
\begin{abstract}

We show that the low energy limit of the minimal supersymmetric 
Left-Right models is the supersymmetric standard model with an exact
R-parity. The theory predicts a number of light Higgs scalars and 
fermions with masses much below the $B-L$ and $SU(2)_R$ breaking scales.
The non-renormalizable version of the theory has a striking prediction 
of light doubly charged supermultiplets which may be accessible to
experiment. 
Whereas in the renormalizable case the scale of parity breaking is
undetermined, in the non-renormalizable one it must be bigger than 
about $10^{10} - 10^{12}$ GeV. The precise nature of the see-saw 
mechanism differs in the two versions, and has important implications 
for neutrino masses.

\end{abstract}

\section{Introduction}

One of the central issues, if not the main one, in the Minimal
Supersymmetric Standard Model (MSSM) is what controls the strength of
R-parity breaking. The suppression of (some or all) R-parity violating
couplings in 
the MSSM is essential to avoid catastrophic proton decay rates, and
determines the fate of the lightest supersymmetric particle (LSP). The
most appealing rationale for an otherwise {\it ad hoc} discrete
symmetry would be to have it as an automatic consequence of a gauge 
principle \cite{m86}.

This is more than an aesthetic issue, for only gauge symmetries are
protected against possible high scale violations such as, for example,
those arising from quantum gravitational effects.
Since in the MSSM the action of R-parity on the  superfields
may be written as $R = (-1)^{3(B-L)}$ \cite{m92}, theories with gauged
$B-L$ may be regarded as the minimal framework to implement this idea
 \cite{m86,fiq89,ir82}.
$B-L$ symmetry is naturally, indeed ineluctably, incorporated in
Left-Right symmetric theories, which provide an understanding of
Parity violation in Nature \cite{ps74,mp75,sm75,s79}. The construction
of a consistent supersymmetric Left-Right theory for generic values of
the Parity breaking scale ($M_R$) thus becomes essential. A
considerable amount of work has been done on theories with  low $M_R$
(that is, $M_R \sim 1 - 10 M_W$ ) regarding the  construction of the theory
\cite{km93,km95}. On the other hand, 
only recently  have there been
attempts to study the more realistic case of $M_R \gg M_W$
\cite{abs97,ams97}. In this paper we provide a systematic study of
minimal supersymmetric Left-Right theories (MSLRM) with an arbitrarily
large scale of Parity breaking and controllable R-parity violation.

This forces us to focus on the version of the theory with the conventional
implementation of the see-saw mechanism \cite{grs79,y79,ms80a}. By this we mean
that the right handed neutrino majorana mass arises at the
renormalizable level.
 However, the following problem arises here:
such a renormalizable theory with the minimal Higgs content 
simply does not allow for any spontaneous symmetry breaking 
whatsoever \cite{c85}. There are two possible ways out of this
impasse: one can either extend the Higgs sector \cite{c85,km95,abs97} 
or allow for nonrenormalizable terms in the 
superpotential \cite{mr96a,ams97}. 

We first concentrate on the renormalizable version of the theory,
which is both  more conventional and simpler to analyze from
the point of view of vacuum structure.  We then apply the same
techniques to the non-renormalizable version, and compare the physical
implications of both models.  This should not imply that we take the
non-renormalizable version less seriously; this is the
minimal theory in terms of the particle spectrum, and it provides the
supersymmetric version of the minimal Left-Right theory.

Although in \cite{abs97,ams97} the vacuum structure of these theories 
was studied, in this paper we present for the first time  a complete
and correct analysis of the lifting of the dangerous D-flat
directions.  Among other things, we learned that unless the sign of
various soft mass terms is positive many of  the flat directions would
not be lifted. This is discussed at length.

Our main conclusion is that unless electromagnetic charge invariance
is violated, R-parity remains unbroken. More precisely, the effective
low-energy theory becomes the MSSM with exact R-parity. This is true
in both versions of the theory. On the other hand, the precise nature
of the see-saw mechanism does depend on whether the symmetry breaking
is achieved through a renormalizable or nonrenormalizable superpotential.

Besides R-parity conservation, another important experimental
signature of these theories is the presence of a number of charged
Higgs supermultiplets whose masses are      
much below $M_R$. For reasonable
choice of parameters, they are expected to lie near the electroweak 
scale. In the
nonrenormalizable version, these light supermultiplets include doubly
charged ones \cite{ams97},  which makes this model specially
interesting from the point of view of experiment. We present in this
paper the complete particle spectrum  for both models.

Another important consequence of our work lies in the possible 
grand unified or superstring extension of left-right symmetry.
Namely, in the literature, one often assumes the extended
survival principle for Higgs supermultiplets. By this one means
that the particles which by symmetries are allowed to be heavy
do indeed become so. The essential lesson of our paper is that this
is completely wrong, since we find a plethora of light Higgs states
which evade the above principle. This conclusion is not new;
it was noticed already in the early papers on the supersymmetric
SO(10) grand unified theory \cite{am83,mp83}. Unfortunately this fact 
is usually overlooked in the literature.

In the next section, we introduce the left-right supersymmetric model
and discuss possible minimal choices for the Higgs sector. We also
summarize the standard method for studying the structure of
supersymmetric vacua, namely the one based on the  characterization of
the  flat
directions of the supersymmetric potential by  holomorphic gauge invariants
of the chiral superfields. In  Section \ref{s3} we apply this method
to analyze the structure of the vacuum of the renormalizable model.
 In Section \ref{s4} we use these results to prove that R-parity (and
therefore both baryon and lepton number) remains an exact symmetry of
the low energy effective theory. We devote Section \ref{s5}  to the study
of the spectrum of the theory, paying special attention to the light
sector. Section \ref{s6} is where the non-renormalizable model is taken up,
and compared to the renormalizable version. Finally, we present our
conclusions and outlook in Section \ref{s7}.

\section{Supersymmetric Left-Right theories}
\label{s2}

The Left-Right symmetric model of gauge interactions treats  fermions of 
 opposite chiralities in a symmetric way by extending the Standard Model 
gauge group to  $SU(3)_c \times 
SU(2)_L \times SU(2)_R \times U(1)_{B-L} $. Thus the anomaly free
global $B-L$ symmetry of the Standard Model is inescapably promoted to
a gauge symmetry in this picture. To obtain left right symmetric
Yukawa interactions that can give rise to the fermion masses it is
necessary to promote the Standard Model Higgs to a bidoublet, and 
realistic fermion mass matrices
require at least two bidoublets in the supersymmetric case.  

In the supersymmetric version of this theory we thus supersymmetrize
the gauge sector in the standard way and introduce  three 
 generations of quark and leptonic chiral superfields with the 
following transformation properties:
\begin{eqnarray}
Q=(3,2,1,1/3)&\;\;\;\;\;  & Q_c=(3^*,1,2,-1/3) \nonumber \\ 
L=(1,2,1,-1)& \;\;\;\;\; & L_c=(1,1,2,1) 
\label{qulep}
\end{eqnarray}
\begin{equation}
 \Phi_i=(1,2,2^*,0) \quad (i = 1, 2)
\end{equation}
where the numbers in the brackets denote the quantum numbers under
$SU(3)_c$, $SU(2)_L$, $ SU(2)_R$ and $ U(1)_{B-L}$ respectively 
(generation indices are understood). In our convention, 
\begin{equation}
L = \left(\begin{array}{c} \nu \\ e \end{array} \right) \quad 
 L_c = \left(\begin{array}{c} \nu_c \\ e_c \end{array} \right)
\end{equation}
so that 
 $L \to U_L L$ under $SU(2)_L$, but $L_c \to U_R^* L_c$ under 
$SU(2)_R$, and similarly for quarks. Also, $\Phi \to U_L \Phi U_R^\dagger$.

The non trivial question that now arises concerns the mechanism for
the spontaneous violation of Left-Right (LR) symmetry,
namely, the selection of a suitable minimal set of Higgs fields to
break this symmetry.  
Furthermore, since the model necessarily includes a right handed neutrino 
a mechanism to explain the observed suppression of neutrino mass (if
any) is also necessary. Indeed one of the most appealing features of
these models is precisely that they provide a natural (``see-saw'')
mechanism to explain this  
suppression. In the non supersymmetric case, of the  two simplest choices for 
the Higgs fields, namely doublets or triplets with respect to
$SU(2)_{L,R}$, only the latter  allows one to realize the above
scenario . However, the inclusion of non-renormalizable operators in
the action can be used to introduce small masses for the neutrino even
in the doublet case \cite{abs92}. 
 
In the supersymmetric case life with doublets is even harder. Since
they carry one unit of $B-L$ charge they are odd under R-parity and
thus the scales of R parity  and LR symmetry breaking must
coincide. Thus the smallness of  
R-parity violation in the doublet case can only be achieved {\it ad 
hoc } as in the  MSSM. Therefore in this paper we choose to work with
triplets. 
 In the concluding section we will discuss the doublet alternative at
greater length. 
The cancellation of $B-L$ anomalies requires the usual doubling of
supermultiplets and thus the minimal choice of Higgs for LR breaking
must include the multiplets  below: 

\begin{eqnarray}
  \Delta = (1,3,1,2)  , \quad 
  \overline{\Delta}  = (1,3,1,-2) \nonumber \\
\Delta_c = (1,1,3,-2), \quad \overline{\Delta}_c = (1,1,3,2)
\label{minimal}
\end{eqnarray}
where $\Delta \to U_L \Delta U_L^\dagger$, but again
$\Delta_c \to U_R^* \Delta_c U_R^T$.
Left-Right symmetry can be implemented in these theories either as a
parity transformation

\begin{eqnarray}
Q              \leftrightarrow     {Q_c}^* ,\quad 
L              \leftrightarrow     {L_c}^* ,\quad
\Phi_i         \leftrightarrow     {\Phi_i}^\dagger,  \nonumber \\
\Delta         \leftrightarrow  {\Delta_c}^* ,\quad
\overline{\Delta}   \leftrightarrow   \overline{\Delta}_c^* \, , 
\end{eqnarray}
or as a charge conjugation

\begin{eqnarray}
Q              \leftrightarrow     Q_c ,\quad 
L              \leftrightarrow     L_c ,\quad
\Phi_i         \leftrightarrow     {\Phi_i}^T,  \nonumber \\
\Delta         \leftrightarrow  \Delta_c ,\quad
\overline{\Delta}   \leftrightarrow   \overline{\Delta}_c \, .
\end{eqnarray} 

The latter definition has an advantage from the point of view
of grand unification, since it is an automatic gauge symmetry in
SO(10). If one is not interested in the nature of CP violation it
makes no difference whatsoever which of the two definitions one uses.
Strictly speaking, we do not even need this discrete symmetry in what
follows, since in the supersymmetric limit all the minima are
degenerate. However, the central challenge in Left-Right theories is
the breaking of parity, so we include it in order to show that it can
be done consistently and in accord with experiment. For the sake of
possible grand unification and transparency of our formula we choose
the latter one.

With this set of multiplets, however, the most general renormalizable
superpotential that one can write for the triplets which are to
accomplish the  breaking of parity is merely

\begin{equation}
 W_{LR}= 
       i {\bf f}( L^T \tau_2 \Delta L    
     + L_c^T\tau_2 \Delta_c L_c ) + 
m_\Delta  ( {\rm  Tr}\, \Delta \overline{\Delta} 
       +   {\rm Tr}\,\Delta_c \overline{\Delta}_c ) \, .
   \label{minsuperpot}
\end{equation}

Since we are considering the case where $M_R \gg M_S$  the SUSY
breaking scale, the minimization of the potential is to be
accomplished by setting the F terms for the chiral superfields and the
D terms for the gauge fields to zero.  Then  
it immediately follows that  the vevs of  $\Delta, \Delta_c$ must vanish 
identically while those of of  $\overline{\Delta}, \overline{\Delta}_c$ are
determined in terms of the vevs of $L,L_c$ respectively. Since the
participation of squark vevs in the symmetry breaking would lead to
charge and color breaking (CCB) 
minima we shall assume that their vanishing is ensured by suitable
soft mass terms. Given that the squark vevs vanish, the form of the
the D term for the $B-L$ gauge field is then 

 \begin{equation}
D_{B-L} = - L^\dagger L  - 2 {\rm Tr}\overline\Delta^{ \dagger}\overline\Delta
 + L_c^\dagger L_c   + 2 {\rm Tr}\overline\Delta_c^\dagger\overline \Delta_c 
\, .
\label{dflat}
\end{equation}

It is clear that vanishing of this D term requires that 
 symmetry breaking in the left and right sectors  must occur 
at the same scale.

To evade this difficulty we introduce an additional set of triplets:
\begin{equation}
\Omega = (1,3,1,0) , \quad \Omega_c = (1,1,3,0)
\end{equation}
where under Left-Right symmetry $\Omega \leftrightarrow \Omega_c$.

The inclusion of this set of multiplets has the additional attraction
of allowing a separation of the scales where parity and $B-L$ symmetry
are broken. 

\section{Symmetry Breaking}
\label{s3}

We next turn to the minimization  of the potential of the
supersymmetric gauge theory introduced in Section \ref{s2}. The most
general gauge invariant superpotential that leads to a renormalizable
action is   
\begin{eqnarray}
 W_{LR}&=& {\bf h}_l^{(i)} L^T \tau_2 \Phi_i \tau_2 L_c 
+ {\bf h}_q^{(i)} Q^T \tau_2 \Phi_i \tau_2 Q_c 
       +i {\bf f} (L^T \tau_2 \Delta L\nonumber \\    
     &  &+ L^{cT}\tau_2 \Delta_c L_c) + 
m_\Delta ( {\rm  Tr}\, \Delta \overline{\Delta} 
       +   {\rm Tr}\,\Delta_c \overline{\Delta}_c ) \nonumber \\
   & & +{m_{\Omega} \over 2} ( {\rm Tr}\,\Omega^2 +
           {\rm Tr}\,\Omega_c^2 )
 + {\mu_{ij} \over 2} {\rm Tr}\,  \tau_2 \Phi^T_i \tau_2 \Phi_j 
        \nonumber \\
  & &     +a ({\rm Tr}\,\Delta \Omega \overline{\Delta} +
        {\rm Tr}\,\Delta_c \Omega_c \overline{\Delta}_c )\nonumber \\
& &  + \alpha_{ij}( {\rm Tr}\, \Omega  \Phi_i \tau_2 \Phi_j^T \tau_2 
       + {\rm Tr}\, \Omega_c  \Phi^T_i \tau_2 \Phi_j\tau_2  )
\label{superpot}
\end{eqnarray}
with 
$\mu_{ij}  =  \mu_{ji} $,
$\alpha_{ij} = -\alpha_{ji}$, ${\bf f}$ and ${\bf h}$ are symmetric
 matrices,  and generation and color indices are understood.

Typically in the minimization of the potential of a SUSY gauge theory
one finds that the space of vacua (the ``moduli space'') may consist
of several sectors  
corresponding to ``flat directions'' running out of various minima 
that would be 
isolated if a suitably smaller set of chiral multiplets had been used.
For example in a $SU(5)$ SUSY gauge theory with  a $\bf{24}$ of
$SU(5)$ as its only chiral multiplet one finds that the permissible
vacua with a renormalizable potential are discrete  (namely the ones
corresponding to the two maximal little groups of $SU(5)$  besides the
trivial one with the full $SU(5)$ unbroken). 
On the other hand, the introduction of additional matter multiplets such as 
a ${\bf{\overline 5 + 10}}$ anomaly free pair quickly leads to a proliferation
of flat directions emerging from these discrete minima.

In what follows we shall use an elegant and powerful method for  
characterizing the vacua of supersymmetric gauge theories 
\cite{bdfs82,ads84,ads85,lt96}.
The essence of this method is simply the following general result: a)
the space of field vevs satisfying the D-flatness conditions
$D_{\alpha}=0$ 
in a supersymmetric gauge theory is coordinatized by the independent
holomorphic gauge invariants that may be formed from the chiral gauge
multiplets in the theory. Further, b) the space of field vevs
satisfying the D and F -flatness conditions is coordinatized by the
holomorphic invariants left  undetermined  
by the imposition of the conditions $F=0$ for each of the chiral
multiplets  in the theory.

The following simple example will serve to clarify the method.
Consider a $U(1)$ gauge theory with two chiral multiplets $\phi_{\pm}$
with gauge charges $\pm 1$. Then the condition $D=0$ requires only
$|\phi_+|=|\phi_-|$. Since gauge invariance can be used to rotate
away one field phase we are left with a magnitude and a phase i.e one
complex degree of freedom left undetermined. Result a) above predicts
this since the only independent holomorphic gauge invariant in this
case is simply $\phi_+\phi_-$ . Now consider the effects of a
superpotential 
$W=m \phi_+\phi_-$. The F flatness condition now ensures that both
vevs vanish so that the D flat manifold shrinks from the complex line
parametrized by $c=\phi_+\phi_-$ to the single point $c=0$.

Thus, in principle, one should proceed by building all the holomorphic
gauge invariants, establish which ones are left undetermined by the
F-flatness conditions and then discuss how the soft SUSY breaking
terms may be used to lift those that are phenomenologically
unacceptable - such as CCB directions. The analysis with the complete
set of fields is however sufficiently complex to  
motivate a simplified approach to the problem.

On phenomenological grounds it is clear that  the bidoublet and squark
fields cannot obtain  vevs at the large scale. Therefore we omit them
from our  analysis of the symmetry breaking at the right handed scale.
Even if they participate in flat directions running out of 
the parity breaking minimum, as long as their
 soft mass terms are taken (as  usual) to be positive their vevs at the
high scale will vanish. On the other hand, since  large vevs for the 
sneutrinos in the right handed sector  are {\it a priori} admissible
the $L$ and $L_c$ fields should be retained in the analysis.

The F-flatness conditions that follow from the the superpotential 
(\ref{superpot}) are as follows:
\begin{eqnarray}
F_{\overline\Delta} &=& m_\Delta \Delta +
 a (\Delta \Omega -{1\over 2} {\rm Tr}\,\Delta\Omega) =0 \nonumber 
 \\
F_{\overline{\Delta}_c} &=& m_\Delta  \Delta_c  
+  a (\Delta_c \Omega_c -{1\over 2} {\rm Tr}\,\Delta_c\Omega_c)=0
\nonumber  \\ 
F_\Delta &=& m_\Delta  \overline\Delta +i {\bf f} L L^T\tau_2 
+ a (\Omega\overline\Delta-{1\over 2}{\rm Tr}\,\Omega\overline\Delta)=0
\nonumber  \\
F_{\Delta_c} &=& m_\Delta \overline\Delta_c +i {\bf f}
  L_c L_c^T \tau_2
+ a (\Omega_c{\overline\Delta}_c-{1\over 2}{\rm Tr}\,
\Omega_c\overline\Delta_c)
=0\nonumber  \\
F_\Omega &=& m_\Omega \Omega
       +a (\overline\Delta\Delta  -{1\over 2} {\rm Tr}\,\overline\Delta\Delta)
= 0 \nonumber \\    
F_{\Omega_c} &=& m_{\Omega} \Omega_c
       +a (\overline\Delta_c\Delta_c  -{1\over 2} {\rm Tr}\,\overline\Delta_c
 \Delta_c)
=0\nonumber \\
F_{L} &=& 2 i {\bf f} \tau_2 \Delta L = 0\nonumber \\
F_{L_c} &=& 2 i {\bf f} \tau_2 \Delta_c L_c = 0
\label{fflat}
\end{eqnarray}

In the above we have self consistently  set the bidoublet and squark  
fields, which must 
have zero vevs at scales $\gg M_R$ to zero.
  
Multiplying the triplet equations by triplet fields and taking traces it 
immediately follows 
that

\begin{eqnarray}
&{\rm Tr}\,\Delta^2 = {\rm Tr}\,\Delta
\Omega = {\rm Tr}\,\overline\Delta
\Omega= 0\nonumber \\
& m_\Delta {\rm Tr}\,\Delta {\overline\Delta} =
m_\Omega  {\rm Tr}\, \Omega^2 =
a {\rm Tr}\,\Delta {\overline\Delta} \Omega 
\nonumber \\
&{\rm Tr} \Delta {\overline \Delta} \left(a^2 {\rm Tr }\Omega^2 - 2
m_\Delta^2 \right) = 0 
\label{fixers} 
\end{eqnarray}
with corresponding equations  {\it mutatis mutandis } in the right
handed sector. 
Thus it is clear that in either sector all three triplets are zero or non 
zero together. By 
choosing the branch where ${\rm Tr}\, \Omega_c^2 = 2 m_\Delta^2/
a^2 $ but ${\rm Tr}\, \Omega^2 =0$ we ensure that the triplet vevs break
$SU(2)_R$ but not $SU(2)_L$.
The field content of the triplets is

\begin{equation}
  \Delta_c   =\left ( \matrix{\delta_c^-/\sqrt{2}
         & \delta_c^{--}\cr
                     \delta_c^0 &-\delta_c^-/\sqrt{2} \cr } \right ),
\quad  \overline\Delta_c   =\left ( \matrix{
  \overline\delta_c^+/\sqrt{2}      & \overline\delta_c^0\cr
  \overline\delta_c^{++} & -\overline\delta_c^+/\sqrt{2} \cr } \right ),
\quad  \Omega_c   =\left ( \matrix{
      \omega_c^0/\sqrt{2}   & \omega_{c\, 1}^- \cr
     \omega_{c\, 2}^+ &-\omega_c^0/\sqrt{2}  \cr } \right )
\label{fields}
\end{equation}
where superscripts denote electromagnetic charges 
\begin{equation}
Q_{em}= T_{3L} + T_{3R} +
{{B-L}\over 2}.
\end{equation}
  
One can use the 3 parameters of the $SU(2)_R$ gauge freedom 
to set the 
diagonal elements of $\Delta_c$ to zero so that it takes the form 
\begin{equation}
\left < \Delta_c \right > =\left ( \matrix{
                      0 &\langle \delta_c^{--}\rangle \cr
                     \langle\delta_c^0\rangle &0 \cr } \right ) \, .
\label{rotation}
\end{equation}
Now (\ref{fixers}) gives $\langle\delta_c^{--}\rangle\langle 
\delta_c^0\rangle = 0$, which implies the electromagnetic
 charge-preserving form for  $\langle \Delta_c \rangle$.
Next it is clear that the Majorana coupling matrix $f_{ab}$ must be 
non-singular if the 
see saw mechanism which keeps the neutrino light is to operate.
 Then it immediately follows from the condition $F_{L_c}=0$, namely, 
\begin{equation}
2 i {\bf f}_{ab}\left ( \matrix{
                      0 & 0 \cr
                     \langle\delta_c^0\rangle &0 \cr } \right )
\left ( \matrix{\nu_c \cr e_c \cr} \right )^b =0
\label{flexplicit}
\end{equation}
 that the sneutrino 
vevs in the 
right-handed sector must vanish. Thus any vev of $L_c$ that appears at the 
high scale 
must necessarily break charge. We ensure that it (together with L) 
vanishes by suitably 
positive soft masses, just as for the squarks. In the Appendix we shall 
exhibit the flat 
directions out of the parity breaking triplet sector vacua associated 
with the slepton fields 
and show that charge is broken in both the left- and right-handed lepton 
sectors along these flat directions.

In the case with triplets alone we first list the $SU(2)_{L,R}$
invariants with their $B-L$ charges. The
gauge invariants can then be generated from these by multiplying
invariants whose charges sum to zero. 
The invariants are :  

\begin{equation}
\begin{array}{c|c}
B-L \,{\rm  charge} & {\rm Invariant}\\
\hline
\; \; 4 & x_1 = {\rm Tr}\,\Delta^2\\
\;\; 2 &  x_2 = {\rm Tr}\, \Omega \Delta \\
\;\; 0 &  x_3 = {\rm Tr}\, \Delta {\overline\Delta} \,,\,\,
x_4 = {\rm Tr}\, \Omega^2\,,\,\,
x_5 = {\rm Tr}\, \Omega \Delta {\overline\Delta}\\ 
-2 &  x_6 = {\rm Tr}\, \Omega \overline\Delta \\
-4 & x_7 = {\rm Tr}\,\overline\Delta^2\\
\end{array}
\label{invariants}
\end{equation}
plus the corresponding invariants $x_i^c$, with opposite charges,
 built from the right-handed fields.
Without the leptons, besides conditions (\ref{fixers}) we have also
\begin{equation}
{\rm Tr}\, \overline\Delta^2 = {\rm Tr}\, \overline\Delta_c^2 =0 \, .
\end{equation}
Notice that this fixes the values of all the $x_i^{(c)}$  and hence, in
fact, all the values of all the gauge invariants that one can form
from the triplet fields. Any flat directions running out of the vacua
allowed by minimizing the potential of the triplets alone (i.e the
trivial and equal left-right vevs vacua which preserve parity and the
2 asymmetric vacua that violate it) must involve the fields we have
omitted from the analysis. If these fields have zero vevs at the high
scale due to positive soft mass terms then the vacua at the high scale
are isolated and, in particular, the parity breaking vacuum
described above is phenomenologically viable. 

It is easy to use the equations (\ref{fixers}) and (\ref{rotation}) to see that
the vevs of $\Omega_c,\Delta_c$ are also fixed to have the charge
preserving form: 

\begin{eqnarray}
 \langle \Omega_c \rangle  = \left (\matrix{ w & 0 \cr
                    0 & -w \cr} \right ), &\quad &
 \langle L_c  \rangle = 0 \nonumber \\
\langle \Delta_c \rangle = 
  \left (\matrix{  0 & 0 \cr
                     d & 0 \cr } \right ), &\quad &
\langle \overline\Delta_c \rangle  =  
  \left (\matrix{  0 & \overline d \cr
                     0 & 0 \cr }\right ) \, .
\label{vev1}
\end{eqnarray} 

In fact using the $B-L$ gauge invariance to fix the relative phase of
 $d$ and ${\overline d}$ one  obtains 

\begin{equation}
 w  =  - {m_\Delta \over a} 
\equiv - M_R ,\quad
d = {\overline d} =  \left({ 2m_\Delta m_\Omega \over
{a}^2}\right)^{1/2} \equiv M_{BL}
\label{vev2}
\end{equation}

Notice an interesting property of (\ref{vev2}). If we wish to have 
$M_R \gg M_{BL}$, 
 we need $m_\Delta\gg m_\Omega$, i.e. a sort of inverse hierarchy 
of the mass scales.
Furthermore, this hierarchy has a highly suggestive geometric form:
$m_\Omega \simeq M_{BL}^2/M_R$. One cannot help speculating that
$m_\Omega$ could be originated by  soft supersymmetry breaking terms. 
Namely, in the absence of
 $m_\Omega$ the superpotential (\ref{superpot}) has a global  $U(1)_R$
 symmetry with the following R-charges
\begin{eqnarray}
\Delta,\overline\Delta,\Delta_c,\overline\Delta_c & \;\;\; 1 \nonumber \\
\Omega & \;\;\;0 \nonumber \\
L, L_c, Q, Q_c & \;\;\; 1/2
\end{eqnarray}

The idea is  very simple. We assume
 the above $U(1)_R$ symmetry which implies $m_\Omega=0$ and which
 will be 
broken by the soft breaking terms \footnote{We thank Gia Dvali for
bringing up this point. For an original application of this idea, see 
\cite{d94}.}.  In the gravity-mediated scenario
 of supersymmetry breaking, it is easy to see that $m_\Omega$ can be 
substituted by the soft supersymmetry breaking terms. In other words

\begin{equation}
m_\Omega \sim m_{3/2} \sim M_W \, .
\end{equation} 
This implies $M_{BL}^2 \simeq M_W M_R$ and thus we have only 
one new scale $M_{BL}$ or $M_R$. Of course, for $M_R \gg M_I \equiv
\sqrt{M_{Pl} M_W}$ (where $M_{Pl}$ is the Planck scale),  
non-renormalizable terms could in principle
induce a bigger value for $m_\Omega \sim M_R^2/M_{Pl}$. We should
stress, though, that neutrino physics strongly suggests $M_R$ to be
smaller or of  order $M_I$.

\section{R-parity Breaking}
\label{s4}

As discussed in the introduction, at scales where supersymmetry is valid, 
the invariance of the action in  minimal  Left-Right symmetric theories
 under R-parity is enforced by  B-L gauge
symmetry and supersymmetry. Thus the only possible source of
R-parity violation at scales $\gg m_S \sim 1 TeV$ is spontaneous: 
when a field with odd $3(B-L)$ develops a vev. 
In the natural and minimal versions
of the Left-Right supersymmetric theory that we have considered
here the only electrically neutral fields that can violate R
parity spontaneously without breaking charge are the sneutrinos
in the two sectors. On the other hand we have seen that the
right handed sneutrino vev is strictly zero at the parity
breaking minimum when working to leading order in the ratio
$m_S/M_R \sim M_W/M_R$. Any R-parity breaking in the right
handed sector thus necessarily involves breaking of charge, at
least at high scales $O(M_R)$. In fact, in the Appendix we show
that the leptonic flat directions running out from the parity
breaking vacuum necessarily violate charge in both left and
right sectors. Thus we regard it as physically well motivated to
assume that soft susy breaking mass terms must be such as to
forbid excursions along the R-parity and charge violating flat
directions, 
just as they must protect color. Thus the effective theory
below the scales $M_R, M_{BL}$ will simply be the MSSM 
(with the addition of some new particle states, see next section) with
R-parity operative. It will therefore also possess an effective global
lepton and baryon number symmetry.

Nevertheless, one may worry that the effects of running of the
coupling constants and masses may be such as to induce vevs for
the sneutrino fields.  Such a thing happens in the MSSM when 
 the Higgs
bidoublet mass squared, although positive at high scales, suffers
large negative corrections due to its strong Yukawa coupling to the top
quark, developing a vev at scales $O(M_W)$ \cite{ir82,acw82}. 
The situation
with sneutrino fields is, however, quite different since none
of the leptonic Yukawa couplings are large. Moreover, even if
the leptonic soft masses were small enough to
be overcome by  renormalization effects in going from
$M_{BL}$ down to the EW scale,  a sneutrino vev
could develop only in the left handed sector since the right
handed sneutrinos have superheavy masses. The global lepton
number symmetry of the effective theory below $M_{BL}$ 
then implies that
a left handed sneutrino vev  would in fact lead
to a Goldstone boson: the (doublet) majoron \cite{am82}.
Such a massless doublet majoron is coupled strongly to the Z 
boson and leads to a large contribution to its width which is
ruled out by experiment.

The one remaining possibility is that the violation of lepton
number by the left sneutrino vev, in combination with the
electroweak vevs, may trigger a vev for the right handed
sneutrino. Such explicit violation of the
effective global Lepton number symmetry would in turn 
give a mass to 
the majoron. If this mass were sufficiently large ($>
M_Z/2$) then the contribution of this state to the Z-width would
be suppressed. To see how $\nu^c$ might get a vev consider the
allowed trilinear soft term in the potential 

\begin{equation}
\Delta V_{soft} = ... + m_S \,L^T\tau_2 \Phi_i \tau_2 L_c + ...
\end{equation}
Once $\nu$ and $\Phi$ develop vevs this implies a linear term in
the potential for the right handed sneutrino which will thus
get an expectation value 

\begin{equation}
\langle \nu^c\rangle ={m_S M_W \langle\nu\rangle \over M_{BL}^2} \, .
\end{equation}
This would lead to effective R-parity and global lepton
number violating terms of the  form $m_\epsilon^2 LH$ where 
\begin{equation}
m_\epsilon^2 = {m_S^2 M_W \langle \nu \rangle \over M_{BL}^2} \, .
\end{equation}
Then the ``Majoron'' would get a mass squared of order
\begin{equation}
m_J^2 \simeq m_{\epsilon}^2 {m_S\over \langle \nu \rangle} 
\simeq {m_S^3 M_W\over M_{BL}^2} \, .
\end{equation}
Thus in order that $m_J$ be large enough to evade the width
bound the scale $M_{BL}$ would have to be $O(m_S)$, which is a
corner of parameter space we do not consider in this paper.
We conclude that within the present scenario the bounds on the Z 
width rule out the possibility of R-parity violation due to
sneutrino vevs. In sum, {\it  the low energy effective theory
of the Minimal Supersymmetric Left Right Model is the MSSM
(with some additional particle states) with
strictly unbroken R-parity, and the LSP is stable}.

\section{Mass Spectrum}
\label{s5}
    
  As we  have seen, the symmetry breaking takes place in  two stages. 
 At a large scale
 $M_R = m_\Delta/a$, $SU(2)_R$ 
is broken down to $U(1)_R$ by the vev of $\Omega_c$.
Later the vevs of $\Delta_c$, $\overline\Delta_c$ are turned on at  
$M_{BL} = \sqrt{2 m_\Delta m_\Omega} /a$, breaking
 $U(1)_R \times U(1)_{B-L}$ 
to $U(1)_Y$. However, a third scale appears in the superpotential,
$m_\Omega = M_{BL}^2/M_R$. Let us now examine the mass spectrum,
and see which scales are involved.

\subsection{Higgs sector}

We begin with the masses of the triplets. The results 
(for $M_R\gg M_{BL}$) are summarized in  Table 1.

\begin{center}

\framebox{\begin{tabular}{l|l}
\hspace{3cm}State & \hspace{1.8cm} Mass \\
\hline 
\hspace{0.3cm}$\delta_c^{++}\, , \,\,\,\,\overline\delta_c^{++}$ & \hspace{0.5cm} $2 a M_R$ \\ 
& \\\hspace{0.3cm} $
\delta_c^+ - {M_{BL} \over \sqrt{2} M_R} \omega_{c\,1}^+$ & \hspace{0.5cm}$
a M_R \left[1 + {1 \over 2}\left({M_{BL} \over M_R}\right)^2 \right]$ \\
& \\\hspace{0.3cm}$
\overline\delta_c^+ - {M_{BL} \over \sqrt{2} M_R} \omega_{c\, 2}^+$ & \hspace{0.5cm}$
a M_R \left[1 + {1 \over 2} \left({M_{BL} \over M_R}\right)^2 \right]$ \\
& \\ \hspace{0.3cm}
$ \left(\omega_{c\,1}^+ +{M_{BL} \over \sqrt{2} M_R}
 \delta_c^+ \right) -\left(\omega_{c\,2}^+ +{M_{BL} \over \sqrt{2} M_R}
\overline \delta_c^+ \right) $
 & \hspace{0.5cm}$
2 g M_R \left[1 +{1 \over 4} \left({M_{BL} \over  M_R}\right)^2 \right]$ \\
\hline
& \\\hspace{0.3cm} $
 \left(\omega_c^0 + {(\delta_c^0 + \overline\delta_c^0)\over
 \sqrt{2}} \right)$  & \hspace{0.5cm}$ a M_{BL} \left(1 + {M_{BL} \over 4 M_R} \right)$\\ 
& \\\hspace{0.3cm} $
 \left(\omega_c^0 - {(\delta_c^0 + \overline\delta_c^0)\over
 \sqrt{2}} \right)$  & \hspace{0.5cm}$ a M_{BL} \left(1 - {M_{BL} \over 4 M_R} \right)$\\
 & \\
\hspace{0.3cm} $
{\rm Re}\,(\delta_c^0 - \overline\delta_c^0)$ & \hspace{0.5cm}$
 2 \sqrt{g^2 + g'^2} M_{BL}$\\
\hline
\hspace{0.3cm}$ \Delta \, , \,\,\,\, \overline\Delta $ & \hspace{0.5cm}$
a M_R $\\
\hspace{0.3cm}$ \Omega $ & \hspace{0.5cm}$ a M_{BL}^2/ 2 M_R  $\\
\hline
\hspace{0.3cm}$H \, , \,\,\,\, \overline H $ & \hspace{0.5cm}$ \sim 0 $\\
\hspace{0.3cm}$H' \, , \,\,\,\, \overline H'$ & \hspace{0.5cm}$ \sim M_R $\\
\end{tabular}}

\vspace{0.5cm}
{\bf Table 1}: Mass spectrum for the  Higgs supermultiplets in the 
renormalizable model.
\end{center}

\vspace{0.5cm}

 As could be expected, almost
all the particles get a mass at the scale $M_R$. 
In the right handed
sector, the doubly charged particles $\delta_c^{++}, \overline\delta_c^{++}$ 
do so, simply  through their
explicit $m_\Delta$
terms in the superpotential. The vev of $\Omega_c$ will contribute
with mass terms for the rest of the charged particles, giving to all
of them a large mass $M_R$. 
However,  the  neutral particles  masses correspond to the (in principle)
 lower scale $M_{BL}$.  The reader can check the
manifestation of the superHiggs mechanism: we have states with masses
equal to the charged ($W_R$) and neutral ($Z_R$) gauge boson masses 

\begin{equation}
M^2(W_R) = 4 g^2 M_R^2 + 2 g^2 M_{BL}^2 \;; \;
\quad M^2(Z_R) = 4 (g^2 + g'^2) M_{BL}^2
\end{equation}

In the left-handed sector, on the other hand, masses come directly
through the explicit terms in the superpotential. $\Delta$ and
$\overline\Delta$ have a large mass of order $M_R$. But the mass of $\Omega$ 
is related to the third mass scale we mentioned above,
$M_{BL}^2/M_R$. 
 This is the most interesting prediction of the model:
a complete $SU(2)_L$ triplet of scalars  and fermions, at a relatively low
mass scale, which could be accessible to future experiments. Notice
that for the
analysis in the previous section to be valid, $m_\Omega \simeq 
M_{BL}^2/M_R$ should not be below the scale of the soft supersymmetry 
breaking terms.
In fact, as we have argued in section \ref{s3}, the natural scale 
for $m_\Omega$ is of order $m_{3/2} \sim M_W$, at least in the 
physically motivated picture with the ratio $M_{BL}^2/M_R$ generated
 dynamically. 
The dependence of these new light states on $M_R$ is noteworthy. Both
$M_{BL}$ and $M_R$ are likely to be large enough to be out of direct
experimental search. However, $M_{BL}$ can be indirectly probed
through the usual see-saw induced neutrino mass  (see below), and thus
improving the experimental limits on new non-MSSM states will actually
set {\it upper} limits on $M_R$. Finally, this indirect probe and the 
direct search for $\Omega$ may provide a crucial test of the consistency 
of the theory.

Once $SU(2)_R$ is broken, the bidoublets $\Phi_1, \Phi_2$ get split into
 four $SU(2)_L$  
doublets, and as usual  one fine-tuning is necessary to  keep one pair of them
  light. Namely when $\Omega_c$ gets 
a vev the mass terms for the bidoublets in
the superpotential becomes

\begin{equation}
W(m_\Phi) = {\mu_{ij} \over 2} {\rm Tr}\,  \tau_2 \Phi^T_i \tau_2 \Phi_j
 + M_R \alpha_{ij} {\rm Tr}\, \tau_3
  \Phi^T_i  \tau_2  \Phi_j \tau_2
\end{equation}
Now, writing the bidoublets in terms of $SU(2)_L$ doublets $H_i, \overline
H_i$ as
\begin{equation}
\Phi_i =
\left(H_i,  \overline H_i \, \right) \equiv \left( \matrix{\phi_i^0 &
\overline \phi_i^+ \cr - \phi_i^- & \overline \phi_i^0}\right )
\end{equation}
the mass terms are seen to correspond to
\begin{equation}
W(m_{H,\overline H}) = \mu_{ij} H_i \overline H_j + \alpha_{ij}M_R  (H_i \overline
H_j - \overline H_i H_j ) 
\end{equation}
 With the fine-tuning condition
\begin{equation}
\mu_{11} \mu_{22} - (\mu_{12}^2 - \alpha_{12}^2 M_R ) \simeq 0
\end{equation}
the bidoublets get split into two heavy left-handed doublets $H',
\overline H'$ with
masses $\sim M_R$, and the two MSSM Higgs doublets $H, \overline H$.

\subsection{Neutrino Mass}

Another distinct prediction of this model is that the see-saw
mechanism takes its canonical
 form. By canonical form we mean (in the  single-generation case)
\begin{equation}
\left(\matrix{0 & m_D \cr
m_D & M \cr }\right)
\label{canonical}
\end{equation}
where $m_D$ is the usual Dirac neutrino mass and $M$ is the large
Majorana mass of the right-handed neutrino.  The mass $M$ is induced
through the vev of $\Delta_c$, and thus $M\sim M_R$. Interestingly
enough,  the form (\ref{canonical}) is hard to achieve in
non-supersymmetric theories, for $\Delta$ in general acquires a small
vev after electroweak breaking. Its origin is a term in the
potential linear in $\Delta$, i.e. of the form $\Delta \Phi^2
\Delta_c$ \cite{ms81}.
In other words, the (1,1) mass element
in general is not zero,
and this has important implications for light
neutrino mass spectrum. Namely, the light neutrino mass is given by 
\begin{equation}
m_\nu = {m_D^2 \over M_R}
\label{seenu}
\end{equation}

In the supersymmetric version we are considering, 
though, the form (\ref{canonical}) is exact up to the
order $1/M_{Pl}$. Simply, at the renormalizable level there are no terms
linear in $\Delta$ in the potential. If one admits non-renormalizable
terms cut-off by the Planck scale, along the lines of \cite{ms81} one
finds  that at electroweak breaking $\Delta$ gets a vev of order
$(M_W^2 M_{BL})/(M_R M_{Pl})$. The (1,1) element in (\ref{canonical}) is
thus suppressed respect to the usual see-saw by $M_{BL}^2/(M_R M_{Pl})$,
and is completely negligible for physics much below the Planck scale.

 \section{Non-renormalizable Model}
\label{s6}

As has been pointed out in \cite{ams97}, it is possible to break parity
 even with just the minimal field content given in 
equations (\ref{qulep}-\ref{minimal}), if one allows
 for non-renormalizable interactions, suppressed by inverse powers of a 
large scale $M$. Including dimension four operators, the most general
 superpotential becomes

\begin{eqnarray}
W_{nr} &=&  m ({\rm Tr}\, \Delta \overline \Delta +
 {\rm Tr}\, \Delta_c {\overline \Delta}_c)
+i {\bf f} ( L^T \tau_2 \Delta L+  L_c^T\tau_2 \Delta_c L_c )
\nonumber \\
& & + {a \over 2 M}\left[  ({\rm Tr}\, \Delta \overline\Delta)^2
 +  ({\rm Tr}\, \Delta_c {\overline\Delta}_c)^2 \right]+ {c \over M}{\rm Tr}\,
 \Delta \overline\Delta {\rm Tr}\, \Delta_c \overline\Delta_c \nonumber \\
& & + {b \over 2 M} \left[ {\rm Tr}\, \Delta^2 {\rm Tr}\, \overline \Delta^2 +
 {\rm Tr}\, \Delta_c^2 {\rm Tr}\, \overline \Delta_c^2 \right] + {1 \over M}
 \left[d_1{\rm Tr}\, \Delta^2 {\rm Tr}\,  \Delta_c^2 +
d_2 {\rm Tr}\, \overline \Delta^2 {\rm Tr}\, \overline \Delta_c^2
\right]  \nonumber \\
& & +{\bf h}_l^i L^T \tau_2 \Phi_i \tau_2 L_c +{\bf h}_q^i Q^T
 \tau_2 \Phi_i \tau_2 Q_c + \mu_{ij}{\rm Tr}\,\tau_2 \Phi_i^T \tau_2
\Phi_j
 \nonumber \\ & &
+ {\lambda_{ijkl}\over  M}{\rm Tr}\,\tau_2 \Phi_i^T \tau_2
\Phi_j{\rm Tr}\,\tau_2 \Phi_k^T \tau_2
\Phi_l + {\alpha_{ij}\over M}({\rm Tr}\, \overline \Delta  \Delta
\Phi_i \tau_2 \Phi_j^T \tau_2   + 
{\rm Tr}\,\overline \Delta_c  \Delta_c  \Phi_i^T \tau_2 \Phi_j \tau_2)
 \nonumber \\
& &+ {\beta_{ij}\over M}{\rm Tr}\,\tau_2 \Phi_i^T \tau_2 \Phi_j [
{\rm Tr}\, \Delta
 {\overline\Delta} + {\rm Tr}\, \Delta_c {\overline\Delta}_c ] \nonumber \\
& &+ 
{\eta_{ij} \over M } {\rm Tr}\,\Phi_i \tau_2 \Delta_c \Phi_j^T\tau_2 \Delta
+
{\overline \eta_{ij} \over M } {\rm Tr}\,\Phi_i \tau_2 \overline\Delta_c
\Phi_j^T \tau_2
 \overline\Delta  \nonumber \\
& & + {{\bf k}_{ql} \over M}Q^T \tau_2 L\,Q_c^T \tau_2 L_c
 + {{\bf k}_{qq} \over M}Q^T \tau_2 Q\,Q_c^T \tau_2 Q_c
+ {{\bf k}_{ll} \over M}L^T \tau_2 L\,L_c^T \tau_2 L_c
\nonumber \\ & &+
{{\bf j}\over M} [Q^T\tau_2 Q\,Q^T\tau_2 L +  
Q_c^T\tau_2 Q_c \,Q_c^T\tau_2 L_c]
\label{nonsuperpot}
\end{eqnarray}

 Of course, not all of the terms above play an equally important role.
If a certain renormalizable interaction is already present in the
potential (as is the case for example with the term $(L_c^\dagger
L_c)^2$), one can safely neglect the small corrections of order $1/M$.
It is only when there are no cubic or quartic couplings that we {\it
must} keep the non-renormalizable ones.

As in the renormalizable case, we assume that the soft 
terms are such as to drive the vevs of the  squark and bidoublet
fields to zero. With this assumption it follows, exactly as in the 
renormalizable case, that for a parity breaking but charge preserving 
vev of $\Delta_c$ the vev of the right handed sneutrinos is necessarily
zero. Thus any R-parity breaking due to $L_c$ getting a vev at scales 
$\gg M_S$
would necessarily break charge. Therefore we assume that
 the soft terms forbid
any vevs for the $L_c$ fields. We are therefore left with the problem of
analyzing the F-flatness conditions for the 4 triplets 
$\Delta, {\overline \Delta},\Delta_c, {\overline \Delta}_c$ to determine whether
or not they admit a charge preserving but parity breaking isolated minimum.

The F-terms for the triplets now read

\begin{eqnarray}
F_\Delta &=&(m  + {a \over M} {\rm Tr}\,\Delta\overline\Delta + 
{c\over M}{\rm Tr}\,\Delta_c \overline\Delta_c)\,\overline \Delta +( {b\over M}
{\rm Tr}\,\overline \Delta^2 + {d_1\over M}{\rm Tr}\,
 \Delta_c^2 )\, \Delta =0 \nonumber  \\ 
F_{\overline \Delta} &=&(m  + {a \over M} {\rm Tr}\,\Delta\overline\Delta + 
{c\over M}{\rm Tr}\,\Delta_c \overline\Delta_c)\, \Delta +( {b\over M}
{\rm Tr}\, \Delta^2 + {d_2\over M}{\rm Tr}\,\overline 
 \Delta_c^2 )\, \overline \Delta =0 \nonumber  \\ 
F_{\overline\Delta_c} &=& (m  + {a \over M} {\rm Tr}\,\Delta_c\overline\Delta_c + 
{c\over M}{\rm Tr}\,\Delta \overline\Delta)\,\Delta_c +( {b^*\over M} {\rm
Tr}\,\Delta_c^2 + {d_2\over M}{\rm Tr}\, \overline\Delta^2 ) 
\, \overline\Delta_c =0 \nonumber  \\ 
F_{\Delta_c} &=&(m + {a \over M} {\rm Tr}\,\Delta_c\overline\Delta_c + 
{c\over M}{\rm Tr}\,\Delta \overline\Delta)\,\overline \Delta_c +( {b\over M}
{\rm Tr}\,\overline \Delta_c^2 + {d_1\over M}{\rm Tr}\,
 \Delta^2 )\, \Delta_c  
=0
\end{eqnarray}

In the renormalizable model with an extra triplet $\Omega$, the left and
 right handed sectors are completely decoupled, and the potential admits
 two discrete minima in each sector.
 The trivial one is chosen by the left-handed sector,
 while the right-handed triplets live in the non-trivial one, which is charge
 and color preserving. In the non-renormalizable case, on the contrary, the 
two sectors are coupled in the F-equations. 
 As we now show, the physically relevant 
minimum, for which the vev of the left-handed triplets vanish and that of 
the right-handed triplets respects charge, does not involve any flat 
direction. This guarantees the stability of the vacuum against the 
presence of  soft terms.

The $SU(2)_{L,R}$ invariants are now just
\begin{equation}
\begin{array}{c|c}
B-L \,{\rm  charge} & {\rm Invariant}\\
\hline
\; \; 4 & y_1 = {\rm Tr}\,\Delta^2\\
\;\; 0 &  y_2 = {\rm Tr}\, \Delta {\overline\Delta} \\
-4 & y_3 = {\rm Tr}\,\overline\Delta^2
\end{array}
\label{noninv}
\end{equation}
together with their right-handed counterparts with opposite $B-L$ charges.
Now, the equations  for $F_\Delta$ and $F_{\overline\Delta}$ give

\begin{equation}
\left[ \left(m  + {a \over M} {\rm Tr}\,\Delta\overline\Delta + 
{c\over M}{\rm Tr}\,\Delta_c \overline\Delta_c \right)^2 -\left({b\over M}
{\rm Tr}\,\overline \Delta^2 + {d_1\over M}{\rm Tr}\, 
 \Delta_c^2 \right)\left({b\over M}
{\rm Tr}\, \Delta^2 + {d_2\over M}{\rm Tr}\,\overline 
 \Delta_c^2 \right) \right] {\rm Tr}\, \Delta\overline\Delta = 0
\label{product}
\end{equation}
and the corresponding equation for the right-handed sector.

Out of the two branches allowed by eqn.(\ref{product}) we focus on the one 
specified by $y_2=0$. With this choice the conditions 
\begin{equation}
(b\,  y_3 +d_1\,  y^c_1)\, y_1 =0 \qquad ; \qquad (b\, y_1 +d_2\, y^c_3)\, y_3 =0
\label{leftkiller}
\end{equation}
follow from $F_\Delta = F_{\overline\Delta}=0$. 
These equations are both satisfied on the branch specified by $y_1=y_2=y_3=0$
and it then follows that the only  gauge invariants
which might  remain undetermined, and therefore allow a
flat direction out of this branch,
are $y_2^c$ and $y_1^c y_3^c$. We emphasize we have not chosen a point
on a flat direction but a branch of solutions of the field
equations specified by the conditions $y_i=0$. 
The equations in the right handed sector are now simply 
\begin{equation}
(m M + (a +b) y_2^c) y_1^c =0
\label{first}
\end{equation}

\begin{equation}
(m M + a y_2^c) y_2^c + b y_1^c y_3^c =0
\label{second}
\end{equation}

The two branches of solutions of eqn.(\ref{first}) are (i) $y_1^c=0$ and
(ii) $y_2^c= -(m M)/(a + b)$ . It easily follows from
eqn.(\ref{second}) that apart from the trivial solution where
all invariants vanish the other possibilities are
\begin{equation}
a) \qquad\qquad y_1^c=y_3^c=0 \qquad ; \qquad y_2^c= - {{m M}\over a}
\end{equation}

\begin{equation}
b) \qquad\qquad y_2^c= -(m M)/(a + b ) \qquad ; \qquad 
 y_1^c y_3^c ={M^2 {m}^2 \over (a + b)^2}
\end{equation}

Solution (a) is equivalent to the one found in the renormalizable case, it
is charge preserving and breaks parity. On the other hand, using
$SU(2)_R$ invariance to put the diagonal elements of 
$\Delta_c$ to zero it immediately follows that solution (b) 
implies breaking of charge.

Thus in this model the triplets get a vev just as in the
renormalizable one, eq. (\ref{vev1}), where
\begin{equation}
d =  {\overline d} =  \sqrt{-{m  M \over a}}
\label{vev3}
\end{equation}

\subsection{Mass spectrum}

It can be seen that the non-renormalizable model has distinct features.
 Symmetry breaking occurs in one stage, at a scale
\begin{equation}
M_R \equiv \sqrt{ - {M m \over a}} \, .
\end{equation}
In the analysis, it has been fundamental to assume $m$ not to be smaller
 than the 
supersymmetry breaking scale. With the large scale $M$ of order $M_{Pl}$,
and $m\stackrel{>}{\sim} 1 TeV$, the parity breaking scale becomes
 $M_R \stackrel{>}{\sim} 10^{11} GeV$. However, the mass spectrum of the 
theory in fact involves two scales. Some of the particles in the Higgs 
multiplets will remain ``light'' after $SU(2)_R$ breaking, with masses 
of order $m \sim M_R^2/M$.  Thus as in the renormalizable case we have a 
very interesting phenomenology involving light charged, doubly charged 
and neutral supermultiplets as shown in Table 2. It is appealing also 
here to invoke soft supersymmetry breaking terms to dynamically generate $m$,
in which case it would be of order of the electroweak scale.

\begin{center}
\framebox{\begin{tabular}{l|l}
\hspace{1cm}State & \hspace{1.3cm} Mass \\
\hline 
\hspace{0.3cm}$\delta_c^+ - \overline\delta_c^+ $ &
 \hspace{0.5cm} $
\sqrt{2} g  M_R$ \\ 
\hspace{0.3cm}${\rm Re}(\delta_c^0 - \overline\delta_c^0)\, $
 &
 \hspace{0.5cm} $ 2 \sqrt{g^2 + g'^2} M_R $ \\  
\hline
\hspace{0.3cm}$\delta_c^0 + \overline\delta_c^0$ &
 \hspace{0.5cm} 
$2 a M_R^2/M$ \\ 
\hspace{0.3cm}$\delta_c^{++}\, , \,\,\,\, \overline\delta_c^{++}$ &
 \hspace{0.5cm} $
2 b M_R^2/M  $\\ 
\hline
\hspace{0.3cm}$\Delta\, , \,\,\,\,\overline\Delta $ & \hspace{0.5cm} $
  (a - c) M_R^2/M $\\
\hline
\hspace{0.3cm}$H \, , \,\,\,\, \overline H $ & \hspace{0.5cm}$ \sim 0 $\\
\hspace{0.3cm}$H' \, , \,\,\,\, \overline H'$ & \hspace{0.5cm}$ \sim 
M_R^2/M $\\
\end{tabular}}

\vspace{0.5cm}
{\bf Table 2}: Mass spectrum for the  Higgs supermultiplets in the 
non-renormalizable model.
\end{center}

Notice again the superHiggs mechanism being operative, since 
supersymmetry is not broken. In this case, only the states that
belong to superHiggs multiplets get a mass of order $M_R$; the rest of
the states can only get  a Planck suppressed mass. 
Of particular interest is the presence of two sets of 
 light doubly-charged scalars and fermions 
($\delta^{++}, \overline\delta^{++}$),
and two full $SU(2)_L$ triplets ($\Delta $ and $\overline\Delta$).
They could have masses as low as the supersymmetry breaking scale.
In contrast to the renormalizable case, the search for these particles
sets a {\it lower} limit on $M_R$.

On the other hand, the bidoublet splitting proceeds in an equivalent way
 as in the renormalizable model, with the higher order interactions 
effectively playing the role of the $\Omega$ field. 
The mass terms for $\Phi_i$ come from 
\begin{eqnarray}
W(m_\Phi) &=&  {\mu_{ij} \over 2} {\rm Tr}\,  \tau_2 \Phi^T_i \tau_2 \Phi_j 
+  {\alpha_{ij} \over M} {\rm Tr}\,  \langle \overline \Delta_c 
\Delta_c \rangle \Phi^T_i \tau_2 \Phi_j \tau_2
\nonumber \\
 & &+ {\beta_{ij} \over M} {\rm Tr}\,  \tau_2 \Phi^T_i \tau_2 \Phi_j 
 {\rm Tr}\,\langle \overline \Delta_c \Delta_c \rangle 
\end{eqnarray}
or
\begin{equation}
W(m_\Phi) = ({\mu_{ij} \over 2} + {M_R^2 \over M} \beta_{ij}) {\rm Tr}\,  \tau_2 \Phi^T_i \tau_2 \Phi_j
 + {M_R^2 \over 2 M} \alpha_{ij} {\rm Tr}\, (1 + \tau_3) \Phi^T_i 
 \tau_2 \Phi_j \tau_2 \, .
\end{equation}

In terms of the $SU(2)$ doublets $H_i,\overline H_i$ we have  now
\begin{equation}
W(m_{H,\overline H}) = \mu'_{ij} H_i \overline H_j + 
 {M_R^2 \over 2 M}\alpha_{ij} (H_i \overline
H_j - \overline H_i H_j ) 
\end{equation}

where $\mu'_{ij} = \mu_{ij} +  M_R^2\, ( 2\beta_{ij} + \alpha_{ij}) / M $.
 The crucial difference with the previous model can be seen immediately:
 one will now fine-tune 
\begin{equation}
\mu'_{11} \mu'_{22} - ({\mu'_{12}}^2 -  {M_R^2 \over 2 M} \alpha_{12}^2
) \simeq 0
\end{equation}
so that the two heavy doublets will not have masses at the large 
scale $M_R$ but at  $m = M_R^2/M  \stackrel{>}{\sim} 1 TeV$. 
Just as in the renormalizable case, it is appealing to have a small
scale $m$ generated softly of order $M_W$. In that case, one 
could imagine the $\mu$ terms being large and being fine-tuned 
among each other. However, this does not work, since there would 
be no splitting of the doublets within the bidoublets, but rather
just splitting of the two bidoublets in the L-R symmetric manner
(one complete bidoublet would remain light, and the other one heavy). 
This in the usual manner would imply the vanishing of quark mixing angles.
In other words $\mu_{ij} \simeq m$, and both generated by the 
soft supersymmetry breaking terms. The dynamical generation of 
mass terms of the theory is a rather appealing and long pursued 
scenario, and furthermore in this case there is then no need
for the fine-tuning of large and independent mass scales.
The clear phenomenological test of this idea is the necessary existence 
of two more weak doublet supermultiplets at the experimentally
accessible energy scale.

Therefore although parity 
is broken at $M_R$ the model does not reduce to the MSSM until much later, 
at the lower scale $m$. This has important consequences for the
solution of the strong
CP problem. 
With four Higgs doublets with masses much  below $M_R$, the
running of the Yukawa matrices  
quickly
generates a sizable strong CP phase  \cite{mr96b}. This
forces any viable solution to the strong CP problem based on parity in
supersymmetric models to have $M_R$ of the order of the weak scale
\cite{mrs97}.

\subsection{Neutrino mass}

Another important difference has to do with neutrino mass. We have
seen that in the renormalizable case the see-saw mechanism takes its
canonical form (\ref{seenu}). Now the situation is completely
different, and resembles the non-supersymmetric case. The
non-renormalizable terms now are essential, since they provide the
interaction terms and determine the scale $M_R$. One finds in the
potential the  relevant terms (written schematically)
\begin{equation}
{m \over M} \Phi^2 \overline\Delta_c \Delta + m^2 \Delta^2 
\end{equation}
which gives a vev for $\Delta$
\begin{equation}
\langle\Delta\rangle ={ \langle\Phi^2\rangle \over \sqrt{m M}} \sim
{M_W^2 \over M_R}
\end{equation}
exactly as in the non-supersymmetric case \cite{ms81}. Once again, the
two models lead to different phenomenological implications.

\section{Discussion and Outlook}
\label{s7}

 In this paper we have offered a complete analysis of symmetry breaking in 
minimal supersymmetric Left-Right models with a scale $M_R$ much above 
the electroweak scale. We were led by the requirement that R-parity is
not introduced ad-hoc and that its breaking be controlled 
without fine-tuning. This rules out the possibility of using doublet
Higgs fields to break parity, and leads naturally to triplets, further 
motivated by the see-saw mechanism.
By minimal, then,  we mean theories with the   
minimal Higgs sector needed to achieve the complete symmetry 
breaking down to $SU(3)_C \times U(1)_{em}$. This in practice means
the following
\begin{itemize}
\item At the renormalizable level, one needs to introduce a new
physical scale associated with $B-L$ breaking.  This is achieved
through the introduction of an additional pair of $B-L$ neutral
triplets. The alternative possibility of utilizing parity-odd singlets
does not work \cite{km95}.
\item If one accepts non-renormalizable terms, the Higgs sector
consists just of the supersymmetric extension of the usual states 
needed to generate  consistent fermion and gauge boson masses.  
\end{itemize}

In both cases, the requirement of minimality meant the exclusion of
further singlet states. The central theoretical result of our analysis
is the proof that the physically acceptable minimum does not lie on a
flat direction. Being an isolated point, with a large barrier
separating it from other (non-physical) minima, there is no danger
that it will not be stable on a cosmological scale. 

The two versions of the theory have in common two extremely important 
characteristics 
\begin{description}
\item[(a)] R-parity never gets broken. 
\item[(b)] The low-energy effective theory, besides the usual MSSM
states,  necessarily contains light charged or doubly-charged
superfield multiplets.
\end{description}

What is different is the nature of the see-saw mechanism and of the
precise spectrum of light states. Whereas in the renormalizable
version the see-saw mechanism takes its canonical form, in the
non-renormalizable case, the situation parallels the one in the
non-supersymmetric Left-Right models.  This in general leads to
different neutrino mass spectra, and is experimentally
distinguishable.

 Other important differences arise in the Higgs mass spectrum, as
displayed in tables 1 and 2. In both cases there are two relevant
physical mass scales above $M_W$. In the renormalizable case they are
$M_{BL}$ and $M_R$ (and we discuss the physically interesting case
 $M_{BL} \ll M_R$). The new light supermultiplet is the 
left-handed Higgs triplet $\Omega$, with a mass of order
$M_{BL}^2/M_R$ which could naturally lie near the weak scale
\cite{abs97}. In particular it does so in the case of a dynamically 
generated ratio of the two new physical scales $M_{BL}$ and $M_R$. In
 other words,  $m_\Omega$ is the result of soft terms, which break
 both supersymmetry and an otherwise automatic continuous R-symmetry.
  The light particles  comprise
 both neutral and single-charged scalars and fermions.

 In the
non-renormalizable  model, the  scales $M_{BL}$ and $M_R$ coincide, 
but there is a new high scale $M_{Pl}$. In this case there is a
plethora of new light states with a mass $M_R^2/M_{Pl}$, which
among other fields include the experimentally very interesting
doubly-charged scalars and fermions \cite{ams97}. 
This is the crucial difference
between the two theories. The doubly charged states are of utmost
interest due to their spectacular experimental signatures. 
In the past there have been a number of papers devoted to the
phenomenological implications of supersymmetric left-right theories
with low $M_R$\cite{hmr94a,hmr94b,hmpr96,cd96,glp96,cr97}, in
which the doubly charged states are discussed. Most of this
analysis carries on to our case.

Light doubly-charged particles continue to
exist even if one adds an arbitrary number of gauge singlets  
\cite{cm97}. In this paper, however, we do not include singlets, for the
whole point of our work has been minimality. The non-renormalizable
version is obviously the minimal supersymmetric Left-Right theory.
However, since renormalizability provides the cornerstone of field
theories, the version with an intermediate $B-L$ scale can also be
considered minimal.

Another, equally important, implication of the existence of the new
light supermultiplets is its impact on the running of the gauge
couplings.  The analysis done in the past often relied on the survival
principle, assuming that all the states which by symmetry are allowed
to be heavy become so. This is manifestly wrong, for as we have shown, there
are a number of light scalars and fermions whose existence defies this
principle. Clearly, a new analysis of  unification is required. It is
not enough to take the result of our paper, for there may be
additional light states which survive the large scale breaking of the
underlying grand unified theory. This has already been noted in
the early works on the SO(10) grand unification of this theory \cite{am83,mp83},
but no running has been performed in these papers. 
In view of this, it is not clear at all that these models
can be successfully unified, but we reserve the final judgment for
the future.

\section{Acknowledgements}

This work was completed during the ``Extended Workshop on Highlights 
in Astroparticle Physics'' held from October 15 to December 15 
at ICTP, and has benefited from discussions with 
its participants, in particular K. Babu and G. Dvali. We wish to thank 
K. Benakli  for original collaboration on supersymmetric Left-Right symmetry.
 C.S.A. and A.M. are grateful to ICTP for hospitality. The work of 
 A.R. and  G.S. is partially supported by EEC grant under the TMR
contract ERBFMRX-CT960090.

\section{Appendix}

In this appendix we analyze the flat directions of the
 superpotential when the 
slepton fields in both sectors are retained in the analysis. Since
doublets must occur bilinearly in all invariants it is convenient to
define  composite singlet ($\sigma_{ab} = - \sigma_{ba}$)
 and triplet fields (  $\overline\Delta'_{ab}=  \overline\Delta'_{ba}$) by

\begin{equation}
L_a {\tilde L}_b = {1\over 2} \sigma_{ab}   +
{1\over 2} {\overline\Delta'_{ab}}
\label{Fierz} 
\end{equation}
and similarly in the right handed sector. They obey Fierzing
constraints like 
\begin{equation}
\sigma_{ab}\sigma_{cd} ={1 \over 2}\sigma_{ad}\sigma_{bc} +
 {1\over 4} Tr ( {\overline\Delta'_{ad}} {\overline\Delta'_{bc}} )
\end{equation}

Then besides the $SU(2)_{L,R}$ invariants listed in
{\ref{invariants}} one can form the additional invariants (not
all independent)

\begin{equation}
\begin{array}{c|c}
B-L \,{\rm  charge} & {\rm Invariant}\\
\hline
-4 & x_8^{abcd}={\rm Tr} \,\overline\Delta'_{ab} \overline\Delta'_{cd}  \\
-4 & x_9^{ab}={\rm Tr} \, \overline\Delta \overline\Delta'_{ab}  \\ 
\;\; 0& x_{10}^{ab}={\rm Tr} \,\Delta  \overline\Delta'_{ab}  \\
-2 & x_{11}^{ab}={\rm Tr} \,\Omega \overline\Delta'_{ab} \\
-2 & x_{12}^{ab}={\rm Tr} \, \Delta \overline\Delta \overline\Delta'_{ab}  \\
\;\; 0& x_{13}^{ab}={\rm Tr} \, \Delta \Omega \overline\Delta'_{ab}  \\  
-4 & x_{14}^{ab}={\rm Tr} \, \overline\Delta \Omega \overline\Delta'_{ab} \\
-2 & x_{15}^{abcd}={\rm Tr} \, \overline\Delta  \overline\Delta'_{ab} \overline\Delta'_{cd}
\\ 
-6 & x_{16}^{abcdef}={\rm Tr} \, \overline\Delta'_{ab}  \overline\Delta'_{cd} 
\overline\Delta'_{ef}   \\ 
-2 & x_{17}^{ab} = \sigma_{ab}  
\end{array}
\end{equation}

   and a similar set in the right handed sector. It is easy to show that
the equations ({\ref{fixers}}) continue to hold in the present case.
Thus one can again choose the parity breaking vacuum by selecting the
solution where all (unprimed) left handed triplets except 
$\overline\Delta$ and $x_1^c,x_2^c,x_6^c$ vanish while  
the vevs of $x_3^c,x_5^c$ are fixed by the solution 
$x_4= 2 m_{\Delta}^2/a^2 $. This leaves us with the
invariants $x_7^{(c)} - x_{17}^{(c)}$ to consider.  In the 
right-handed sector, as we saw in Section \ref{s3}, the equation for
$F_{L_c}$ ensures that the right handed sneutrino vevs 
necessarily vanish. Since   
$x_{17}^{(c)}=-i \nu^c_{[a} e^c_{b]}$,  
it vanishes. Due to  the Fierz relations eqn({\ref{Fierz})
 (and a similar one relating the $x_{16}^{(c)}$ to products 
of $x_{17}^{(c)}$ and  $x_8^{(c)}$ ) $x_8^{(c)}$ and  $x_{16}^{(c)}$
 all vanish while 
 
 \begin{equation}
\overline\Delta_{(c){ab}}' = e^c_a e^c_b \tau_+
\label{dlec}
\end{equation}
Then using the F equations in the right handed sector it is easy
to convince oneself that all invariants are fixed in terms of
$x_9^{(c)ab}$. However, due to eqn(\ref{dlec})  only three are
of these are independent and these can be taken to be $x_9^{(c) aa}$.

In the left handed sector one finds that $\overline\Delta$ is
determined  in terms of $\overline\Delta_{ab}'$ by the equation for 
$F_{\Delta}$, thus  only invariants involving $\sigma_{ab} $ or 
$\overline\Delta_{(c)ab}'$ are left. But since all of these 
may be written in terms of products of $x_{17}^{[ab]}$, its is
clear that these are the only independent $SU(2)_L$ invariants
left undetermined in the left sector and are 3 in number.  
Since $x_{17}^{ab}$ carries $B-L$ charge (-2) and 
$x_9^{(c) aa}$ carries (-4) it is clear that one can form 
9 independent gauge invariants which are left undetermined after
imposition of the F constraints:
\begin{equation} 
z_{[ab] d}= x_{17}^{ab} [x_9^{(c) dd}]^{1/2}
\end{equation}

 Thus the manifold of flat
directions running out of the parity breaking vacuum is
parametrized by these 9 complex coordinates. From eqn(\ref{dlec})
and $\sigma_{ab}=-i \nu_{[a} e_{b]}$, it is thus clear that 
the coordinates $z_{[ab] d}$ all involve a product of 
selectron and anti-selectron vevs,
and hence these flat directions violate charge.

\end{document}